\documentclass[aps,prl,showkeys,floatfix,twocolumn,bibnotes,nofootinbib]{revtex4-2}
\usepackage{graphicx,amsmath,amssymb,hyperref,natbib,slashed}
\usepackage[dvipsnames]{xcolor}
\usepackage{xspace}

\usepackage[utf8]{inputenc}

\newcommand{\ds}{{\sf DarkSUSY}}

\newcommand{\di}{\text{d}}

\newcommand{\overbar}[1]{\overline{#1}}

\newcommand{\be}{\begin{equation}}
\newcommand{\ee}{\end{equation}}
\newcommand{\bea}{\begin{eqnarray}}
\newcommand{\eea}{\end{eqnarray}}

\begin{document}
\hfill DESY 20-113

\title{Precise dark matter relic abundance in decoupled sectors}

\newcommand{\AddrOslo}{%
Department of Physics, University of Oslo, Box 1048, N-0316 Oslo, Norway}
\newcommand{\AddrDESY}{%
Deutsches Elektronen-Synchrotron DESY,  Notkestra\ss e 85, D-22607 Hamburg, Germany}

 \author{Torsten Bringmann}
 \email{torsten.bringmann@fys.uio.no}
 \affiliation{\AddrOslo}

  \author{Paul Frederik Depta}
 \email{frederik.depta@desy.de}
 \affiliation{\AddrDESY}

  \author{Marco Hufnagel}
 \email{marco.hufnagel@desy.de}
 \affiliation{\AddrDESY}

 \author{Kai Schmidt-Hoberg}
 \email{kai.schmidt-hoberg@desy.de}
 \affiliation{\AddrDESY}


\begin{abstract}
Dark matter (DM) as a thermal relic of the primordial plasma is increasingly pressured by direct and
indirect searches, while the same production mechanism in a decoupled sector is much
less constrained. We extend the standard treatment of the freeze-out process to such scenarios
and perform precision calculations of the 
annihilation cross-section required
to match the observed DM abundance.
We demonstrate that the difference to the canonical value of this `thermal cross-section'
is generally sizeable,
and can reach orders of magnitude.
Our results directly impact the interpretation of DM searches in hidden sector scenarios.
\end{abstract}

\maketitle

\paragraph*{Introduction.---}%

Cosmological observations require the existence of a dark matter (DM) component that makes up about 80\,\%
of the matter in our Universe~\cite{Aghanim:2018eyx} and likely consists of a new type of elementary
particle~\cite{Jungman:1995df,Bertone:2004pz}. The most often adopted paradigm for DM production
is via freeze-out from the primordial plasma of standard model (SM) particles~\cite{Lee:1977ua}.
This roughly requires weak-scale couplings for DM masses at the electroweak scale -- which has been argued to
be an intriguing coincidence in view of proposed solutions to the hierarchy problem of the SM~\cite{Ellis:1983ew}
-- but the same mechanism also works for lighter DM and correspondingly weaker couplings~\cite{Feng:2008ya}.
The formalism to calculate the thermal relic abundance in these scenarios~\cite{Gondolo:1990dk,Edsjo:1997bg}
is well established and successfully used in a plethora of applications,  {\it e.g.}~for benchmarking
the reach of experimental searches for non-gravitational DM interactions~\cite{Ackermann:2015zua,
Aghanim:2018eyx,Acciari:2020pno,Ahnen:2017pqx,Morselli:2017ree,Mukherjee:2018wke}.
Based on this standard prescription,
several public numerical codes~\cite{Bringmann:2018lay,Ambrogi:2018jqj,Belanger:2006is,Arbey:2009gu}
provide precision calculations of the DM abundance, matching the percent level observational accuracy.

More recently, the focus has shifted to models where DM couples more strongly to particles
in a `secluded' dark sector (DS) than to the 
SM~\cite{Pospelov:2007mp,Feng:2008mu,Pospelov:2008zw,Sigurdson:2009uz,Cheung:2010gj}. This
development is partially motivated by the fact that more traditional DM candidates are increasingly pressured
by the absence of undisputed signals in direct searches as well as at
colliders~\cite{Arcadi:2017kky,Bertone:2018krk,Beacham:2019nyx}, but also from a theoretical
perspective there is no need for sizeable inter-sector couplings.
Remarkably, thermal freeze-out works equally well also in these models, providing a
compelling potential explanation for the observed DM abundance.
As a consequence, couplings needed to achieve this goal are often either implicitly fixed or explicitly targeted 
in various searches for hidden sector 
particles~\cite{Pospelov:2008jd,Mardon:2009rc,Schmidt-Hoberg:2013hba,Tulin:2013teo,Alekhin:2015byh,
Hufnagel:2017dgo,Kahlhoefer:2017umn,Bondarenko:2019vrb}.

Despite this development, relic density calculations in such scenarios have not yet reached the same level of
refinement as for thermal freeze-out in the visible sector. As a critical first step towards bridging this gap, we 
perform here a concise and comprehensive analysis of model-independent aspects of such calculations, 
matching both in spirit and precision the widely adopted treatment of  `standard' freeze-out in the 
visible sector~\cite{Steigman:2012nb}. 

In the first part of this {\it Letter} we discuss (effectively) massless 
annihilation products. Since in this case the main differences to the standard analysis have largely
been identified previously~\cite{Feng:2008mu,DAgnolo:2015ujb,Pappadopulo:2016pkp,Farina:2016llk,Berlin:2016gtr,
Kamada:2017gfc,Kamada:2018hte,Arcadi:2019oxh,Dondi:2019olm,Hambye:2020lvy,Baldes:2017gzw}, our focus 
here is on a pedagogic and easily accessible summary, including the presentation of new benchmark `thermal' 
cross-sections that
can directly be compared to the corresponding visible sector results~\cite{Steigman:2012nb}.
In the second, and main, part of this {\it Letter} we then perform a detailed analysis of 
DM annihilating into DS particles similar in mass. In this case comoving conservation of the total number 
density necessitates an accurate treatment of chemical potentials of {\it all} involved particles, both before and 
during freeze-out, which we provide here for the first time.

\smallskip
\paragraph*{Standard freeze-out.---}%
We start by briefly revisiting the canonical approach.
The number density $n_i$ of DM particles $i=\chi,\bar\chi$ initially in thermal equilibrium with the
SM heat bath at temperature $T$ can be described by the Boltzmann equation~\cite{Gondolo:1990dk}
\be
\label{boltzn_standard}
\frac{\di n_i}{\di t}+3Hn_i= \langle \sigma v\rangle\left(n_{\chi,{\rm eq}}n_{\bar\chi,{\rm eq}} -n_\chi n_{\bar\chi}\right)\,,
\ee
where $H$ is the Hubble rate,
\be
\label{therm_av_SM}
 \left\langle \sigma v\right\rangle
  =  \int_1^\infty\!\!\!\! \di \tilde s\, \sigma v_{\rm lab}\,
 {  2x\sqrt{\tilde s\!-\!1}(2\tilde s\!-\!1)K_1\!\!\left(2{\sqrt{\tilde s}} x\right)}
 /{{K_2}^{2}(x)}\,,
\ee
and $n_{\chi,{\rm eq}}=n_{\bar\chi,{\rm eq}}= g_\chi m_\chi^3 K_2(x)/(2\pi^2 x)$.
Here, $x\equiv m_\chi/T$, $K_j$ are modified Bessel functions of order $j$,
$g_\chi$ denotes the internal degrees of freedom (d.o.f.) of $\chi$,
$\sigma$ is the total cross-section for
DM annihilations, for a center-of-mass energy $\sqrt{s}\equiv2m_\chi \sqrt{\tilde s}$,
and $v_{\rm lab}$ is the velocity of one of the DM particles in the rest-frame of the other.

Let us stress two main assumptions that enter in this widely used form of the Boltzmann
equation. The first is that the DM phase-space distribution is of the form
$f_i\propto f_{\chi,{\rm eq}} =\exp(-E_\chi/T)$, 
i.e.~that the freeze-out happens for $m_\chi\gg T$ and well before kinetic decoupling 
(see Ref.~\cite{Binder:2017rgn} for a treatment of early kinetic decoupling).
The second assumption is that the annihilation products constitute a
{\it heat bath}, in the sense that none of them builds up significant chemical potentials.
Crucially, both assumptions can be violated in decoupled sectors.

\begin{figure}[t]
\includegraphics[width=0.95\columnwidth]{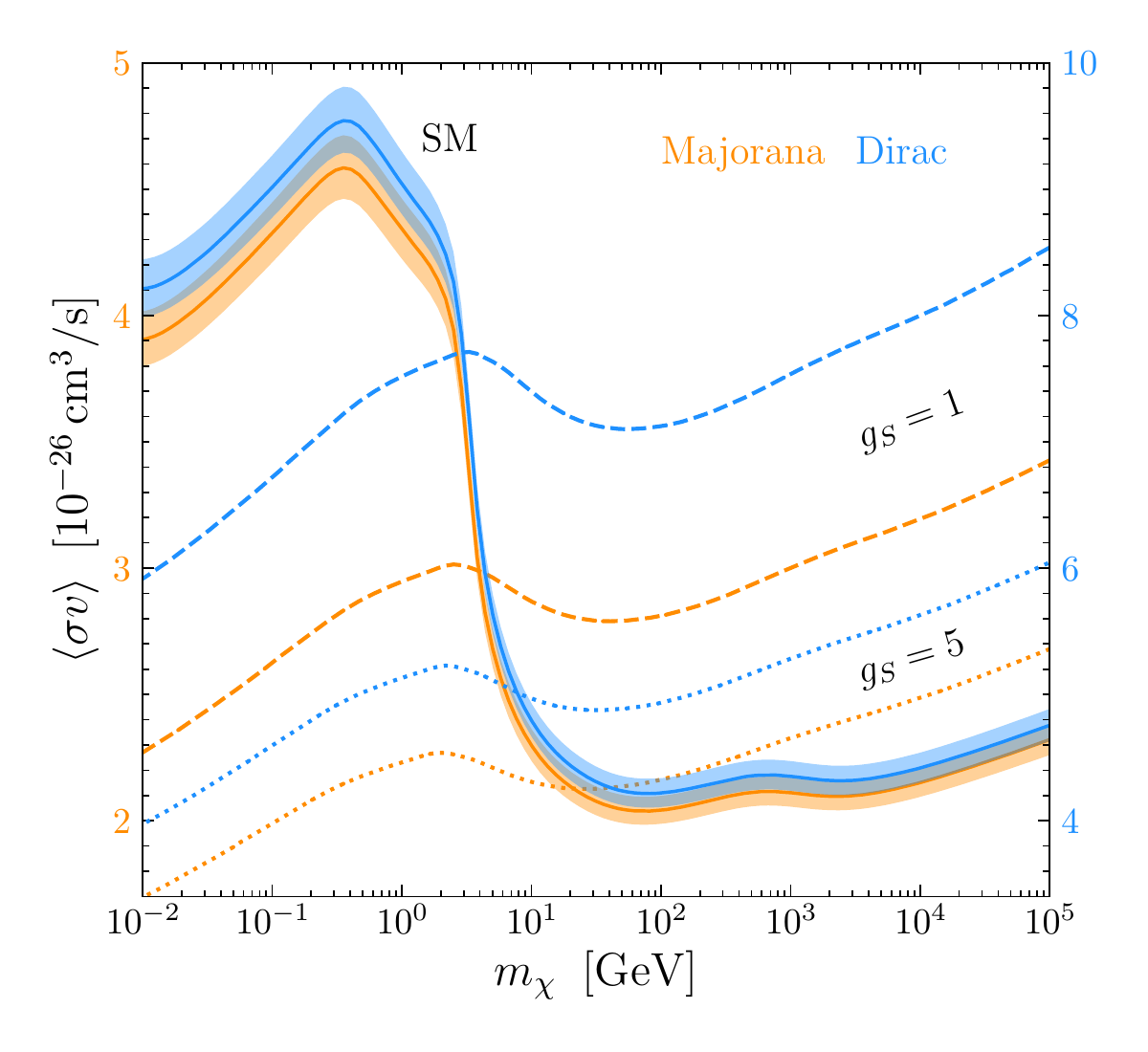}
\vspace*{-2ex}
\caption{The value of a constant thermally averaged annihilation cross-section, $\langle\sigma v\rangle$,
resulting in a relic density of Majorana (orange) or Dirac (blue) particles matching the
observed cosmological DM abundance. Solid lines show the case of DM in equilibrium
with the SM until freeze-out (shaded areas indicate the effect of varying $\Omega_{\rm DM} h^2$
within $3\sigma$~\cite{Aghanim:2018eyx}). Dashed (dotted) lines show the case of DM in equilibrium with
a hidden sector containing $g_S=1$ ($g_S=5$) light scalar degrees of freedom (with $\mu_S=0$),
which decoupled from the SM at $T\gg \max[m_\chi,m_t]$.
Corresponding results for $p$-wave annihilation are presented in Appendix A.
}
\label{fig:no_mu_dark}
\end{figure}

In Fig.~\ref{fig:no_mu_dark} we indicate with solid lines the value of
$\left\langle \sigma v\right\rangle$ in the standard scenario (assuming a constant value of 
this quantity around chemical decoupling)
that is needed to obtain a relic density matching the observed cosmological
DM abundance of $\Omega_{\rm DM}h^2=0.12$~\cite{Aghanim:2018eyx}. The orange solid lines
show the case of Majorana DM (with $g_\chi=2$ and $\Omega_\chi=\Omega_{\bar\chi}=\Omega_{\rm DM}$),
updating the conventionally quoted `thermal relic cross-section' in Ref.~\cite{Steigman:2012nb} with a more
recent measurement of $\Omega_{\rm DM}$ and recent lattice QCD results for the evolution of SM
d.o.f.~in the early Universe~\cite{Drees:2015exa}. 
For comparison, the blue lines indicate the case of Dirac DM
($g_\chi=g_{\bar \chi}=2$ and $\Omega_\chi=\Omega_{\bar\chi}=\Omega_{\rm DM}/2$)
to stress the not typically appreciated fact that the required value of $\left\langle \sigma v\right\rangle$
is {\it not} exactly twice as large as in the Majorana case.

\smallskip
\paragraph*{A secluded dark sector.---}%
The idea~\cite{Pospelov:2007mp,Pospelov:2008zw,Feng:2008mu,Pospelov:2008jd,Sigurdson:2009uz,Cheung:2010gj} 
that DM could be interacting only relatively weakly with the SM,
but much more strongly with itself or other particles in a secluded DS, has received significant
attention~\cite{ArkaniHamed:2008qn,Tulin:2013teo,Ackerman:mha,Batell:2009yf,Alexander:2016aln,
Bringmann:2016din,Pappadopulo:2016pkp,Bringmann:2018jpr}. In such scenarios, both sectors may well have 
been in thermal contact at high temperatures, until they decoupled at a temperature $T_{\rm dec}$. 
The separate conservation of entropy in both sectors then implies a non-trivial evolution of the temperature ratio,
\be
\label{xi_eq}
 \xi(T)\equiv \frac{T_\chi(T)}{T}=
 \frac{\left[{g_*^\mathrm{SM}}(T)/ {g_*^\mathrm{SM}}(T_\mathrm{dec}) \right]^\frac13}
 {\left[{g_*^\mathrm{DS}(T)}/{g_*^\mathrm{DS}(T_\mathrm{dec})}\right]^\frac13}\,,
 \ee
where $g_*^\mathrm{SM,DS}$ denotes the effective number of relativistic entropy d.o.f.~in the two 
sectors. Let us stress that this commonly used relation tacitly assumes that DM is in full equilibrium with 
at least one species $S$ with vanishing chemical potential, $\mu_S=0$ (implying 
$\mu_\chi=-\mu_{\bar\chi}$ as long as DM is in chemical equilibrium).

For a precise description of the freeze-out process of $\chi$ in such a situation
the standard Boltzmann equation \eqref{boltzn_standard} 
needs to be adapted at three places: both {\it i)} the equilibrium density $n_{\rm eq}$ and {\it ii)} the thermal
average $\langle \sigma v\rangle$ must be evaluated at $T_\chi$ rather than the SM temperature $T$,
and {\it iii)} the Hubble rate must be increased to take into account the energy content
of the DS. 
During radiation domination, in particular, this means that
$H^2=(8\pi^3/90)g_{\rm eff}M_{\rm Pl}^{-2}T^4$, where
$g_{\rm eff}\simeq g_{\rm SM} + (\sum_b g_b+\frac78\sum_f g_f)\xi^4$ and the sums runs
over the internal d.o.f.~of all fully relativistic DS bosons ($b$) and DS fermions ($f$)
(in our numerical treatment, we always use the {\it full} expression for $g_{\rm eff}$).
We note that existing relic density calculations for decoupled DSs very often
only take into account a subset of these effects, or implement them in a simplified, 
not fully self-consistent way. 

\smallskip
\paragraph*{Model setup.---}%
Let us for concreteness consider a simple setup where the DS consists of massive fermions $\chi$, acting as DM,
and massless scalars $S$ with $\mu_S=0$, constituting the heat bath.
We assume that the DS decoupled from the SM at
high temperatures, such that ${g_*^\mathrm{SM}}(T_\mathrm{dec})=106.75$ and
$g_*^\mathrm{DS}(T_\mathrm{dec})=g_S+(7/4)N_\chi$ in Eq.~\eqref{xi_eq}, where $N_\chi=1$ ($2$)
for Majorana (Dirac) DM. In Fig.~\ref{fig:no_mu_dark} we show
the `thermal' annihilation cross-section for $\chi\bar \chi\to SS$ in such a scenario,
for different values of $g_S$.
The fact that this differs significantly from the standard case
illustrates the importance of including the effects outlined above in a consistent way.
In this sense, the updated {\it procedure} for relic density calculations directly impacts
a large number of DS models where annihilation also proceeds via an $s$-wave~\cite{Pospelov:2008jd,
Pospelov:2008zw,ArkaniHamed:2008qn,Feng:2009mn, Aarssen:2012fx,Tulin:2013teo, Kaplinghat:2013yxa,Cirelli:2016rnw}
-- even though $\sigma v$ is often velocity-dependent in these cases, impeding 
a literal interpretation of the curves shown in  Fig.~\ref{fig:no_mu_dark}. 
In order to facilitate the study of such more realistic scenarios,
we have updated the general-purpose relic density routines of \ds~\cite{Bringmann:2018lay} 
to perform precision calculations 
of DS freeze-out that self-consistently take into account all three effects discussed above. This allows to 
consider a broad range of relevant models with in principle arbitrary amplitudes, including $p$-wave annihilation
(see also Appendix A) and, {\it e.g.}, Sommerfeld enhancement.\footnote{These updates have been included in release {\sf  6.2.3} of \ds. 
See {\tt https://darksusy.hepforge.org} 
for explicit examples and further details.}

To understand the behavior of the curves in Fig.~\ref{fig:no_mu_dark}, 
we first note that a constant $\langle \sigma v\rangle$ (as in this specific example) is of course 
not affected by a change in $\xi$.
For $g_S=1$, furthermore, the change in $g_{\rm eff}$ and hence the Hubble rate has only a subdominant effect
(but becomes somewhat more important for $g_S=5$).
The main effect visible in the figure thus
originates from changing $n_{\chi,{\rm eq}}(x)\to n_{\chi,{\rm eq}}(x/\xi)$.
For large DM masses and hence freeze-out temperatures, in particular, the heating in the DS due to
$\chi\bar\chi\to SS$, cf.~the nominator of Eq.~\eqref{xi_eq}, 
is more efficient than the heating in the SM,
leading to $\xi>1$ around freeze-out. This leads to a larger DM density, at a given SM temperature $T$,
which has to be compensated for by a {\it larger} $\langle \sigma v\rangle$  to match the observed relic abundance.
Below DM masses of a few GeV, the drop in the SM d.o.f.~until freeze-out is more significant than that in the
DS (especially during the QCD phase transition), leading to $\xi<1$ and hence the need for a {\it smaller} value of 
$\langle \sigma v\rangle$ compared to the standard case represented by the solid lines. 
We finally note that the energy density of $S$ at late times is independent of $m_\chi$.
Expressing it in terms of an effective number of
relativistic neutrino species, this corresponds to $\Delta N_\mathrm{eff} = 0.104 (0.202)$ for Majorana DM
with $g_S=1 (5)$, and $\Delta N_\mathrm{eff} = 0.201 (0.275)$ for Dirac DM -- which is below
current CMB bounds on this quantity, $\Delta N_\mathrm{eff} < 0.29$ (95\% C.L.)~\cite{Aghanim:2018eyx},
but within reach of next-generation CMB experiments~\cite{Ade:2018sbj,Abazajian:2019eic}.

\smallskip
\paragraph*{Chemical potentials during freeze-out.---}%
The above treatment still assumes that the annihilation products are in chemical equilibrium with themselves
during the entire chemical decoupling process of DM. This is consistent for
massless DS particles $S$, where
interactions such as $\chi\bar\chi \rightarrow \chi\bar\chi S$ (or number-changing reactions purely 
within the $S$ sector) will always 
enforce $\mu_S=0$.
For a fully decoupled DS only containing massive degrees of freedom, however,  this is no longer necessarily the case. 
Largely independent of the concrete model realization, in particular,  number-changing interactions of massive 
particles $S$ will cease to be efficient at the latest when the DS temperature drops below their mass, 
$T_\chi\lesssim m_S$. This implies that {\it all} DS particles will generally build up chemical potentials before and
during the freeze-out process (and not only the DM particles, as in the standard scenario).

\begin{figure}[t]
\includegraphics[width=0.95\columnwidth]{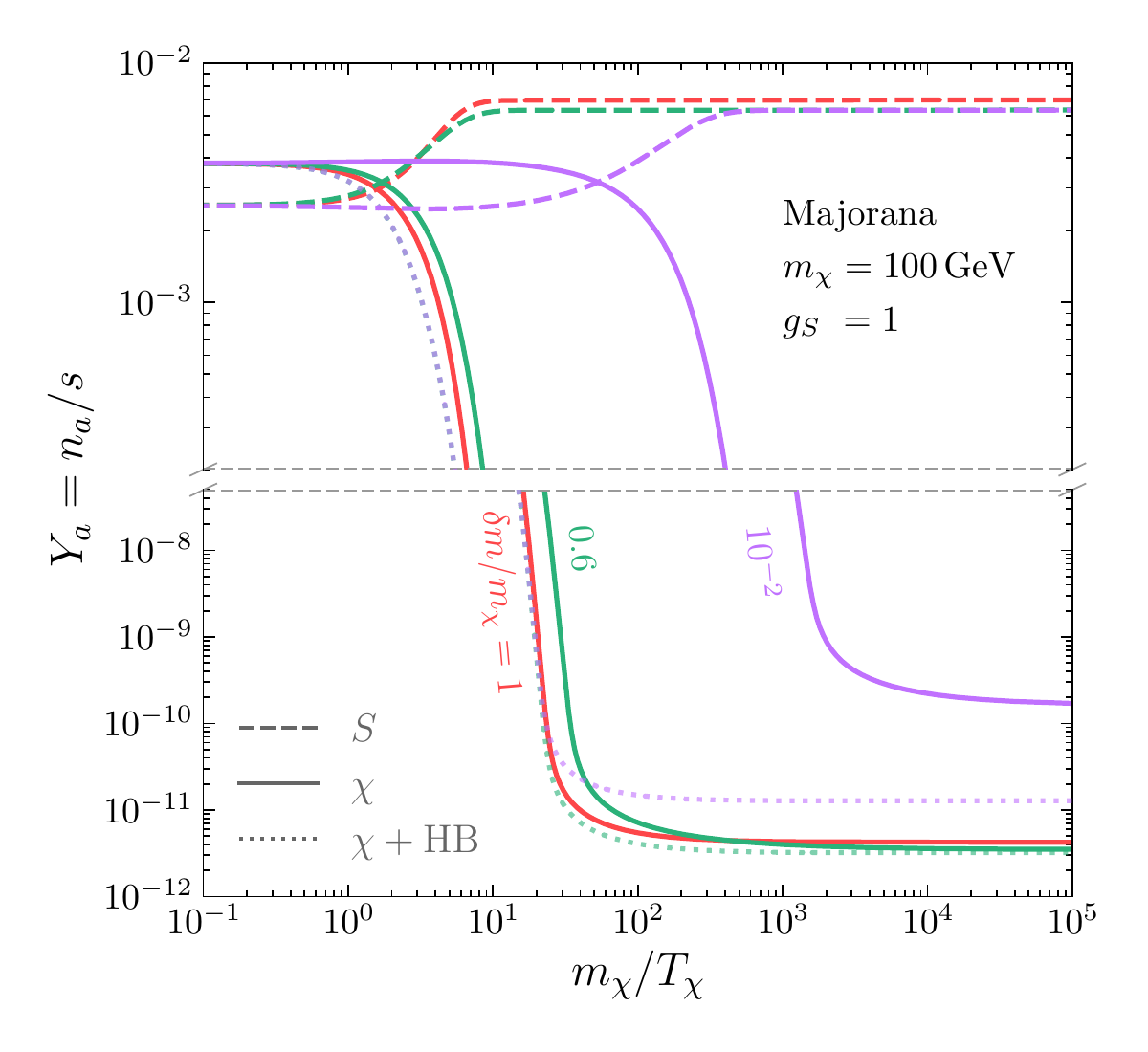}
\caption{%
Evolution of particle abundances $Y_a$ for $\chi$ (solid lines) and $S$ (dashed lines), as
a function of $x/\xi=m_\chi/T_\chi$, for different mass ratios $\delta m/m_\chi\equiv(m_\chi\!-\!m_S)/m_\chi=(1,0.6,10^{-2})$.
For comparison, dotted lines indicate how the DM abundance $Y_\chi$ would evolve when instead using the
standard Boltzmann equation~\eqref{boltzn_standard} assuming thermal equilibrium of $S$ with an additional massless 
DS heat bath particle. All curves are based on the same 
$\left|\overbar{\mathcal{M}}_{\chi\bar{\chi}\rightarrow SS} \right|^2=const.$, adjusted to give the correct relic
density in the limit $m_S\to0$.
}
\label{fig:Yx}
\end{figure}

Let us for illustration consider the same setup as before, but mostly focus on DS particles $\chi$ and $S$ 
close in mass. 
In chemical equilibrium, $\chi\bar\chi\leftrightarrow SS$ then enforces $\mu_\chi+\mu_{\bar\chi}=2\mu_S$.
At $T_\chi\gg m_i$, as argued above, one furthermore has $\mu_S=0$; this implies
$\mu_\chi=-\mu_{\bar\chi}=0$ (where the last equality assumes a vanishing asymmetric DM component). 
When $S$ later develops a non-vanishing chemical potential, on the other hand, that initial condition leads to 
$\mu_\chi = \mu_{\bar{\chi}} = \mu_S > 0$. 
Subsequently, DM decouples chemically from $S$ --  but will typically remain in {\it kinetic} 
equilibrium at least until the end of the freeze-out process. 
The phase-space densities of all 
DS particles are then still given by Fermi-Dirac or Bose-Einstein distributions with temperature
$T_\chi$ and chemical potentials $\mu_\chi = \mu_{\bar{\chi}}\neq\mu_S$~\cite{Kolb:1990vq}. 
We recall that Eq.~\eqref{boltzn_standard} describes the evolution of $n_\chi$ if $\mu_S=0$ and the effect
of quantum statistics can be neglected, i.e.~$f_\chi\propto \exp(-E_\chi/T)$.
In the following we will instead demonstrate how to accurately determine the evolution of 
$n_\chi$ without these two assumptions.

Let us start, for simplicity, with the case where $\chi\bar \chi \leftrightarrow SS$ is the only relevant 
$S$-number changing reaction, implying, e.g., that $S$ is sufficiently long-lived 
(below, we will modify this assumption).
We then  consider the Boltzmann equations for the number densities,
\begin{align}
\dot{n}_i + 3Hn_i = \mathfrak{C}\,,\quad\dot{n}_S + 3Hn_S = - N_\chi \mathfrak{C}\,,
\label{eq:boltzmann_similar}
\end{align}
where $\mathfrak{C}$ is the integrated collision operator in its standard form~\cite{Kolb:1990vq},
as well as energy conservation in the DS during freeze-out, $\nabla_\mu T^{0\mu}_{\rm DS}=0$. 
The latter takes the form
\begin{align}
\dot{\rho}_{\rm DS} + 3H\big[ \rho_{\rm DS} + P_{\rm DS} \big] = 0\,,
\label{eq:friedmann_similar}
\end{align}
with total energy density $\rho_{\rm DS} \equiv N_\chi \rho_{\chi} + \rho_S$ and pressure
$P_{\rm DS} \equiv N_\chi P_{\chi} + P_S$. 
Finally, as long as $\chi$ and $S$ stay in kinetic equilibrium,
all cosmological quantities $Q \in \{ n_{a}, \rho_a, P_a \,|\, a \in \{ \chi, \bar\chi, S\} \}$ can be interpreted as
functions of $T_\chi$, $\mu_\chi$ and $\mu_S$ only. We therefore can use
\begin{align}
\dot{Q} = \frac{\partial Q}{\partial T_\chi}\dot{T}_\chi  +
\frac{\partial Q}{\partial \mu_\chi}\dot{\mu}_\chi + \frac{\partial Q}{\partial \mu_S}\dot{\mu}_S
\end{align}
to transform Eqs.~\eqref{eq:boltzmann_similar} and \eqref{eq:friedmann_similar} into a set of
differential equations for $T_\chi$, $\mu_\chi$ and $\mu_S$, which we solve numerically (with
$\mu_\chi=\mu_{\bar\chi}=\mu_S=0$ as initial condition). 
Note that
Eqs.~\eqref{eq:boltzmann_similar} and~\eqref{eq:friedmann_similar} replace Eqs.~\eqref{boltzn_standard} and~\eqref{xi_eq},
and generally only imply entropy conservation during chemical equilibrium. For the specific 
benchmarks below, chemical decoupling occurs only after significant Boltzmann suppression of $\chi$
and $\bar{\chi}$, 
such that the respective change in total DS entropy can still be neglected.

In Fig.~\ref{fig:Yx} we demonstrate the resulting evolution of the particle abundances $Y\equiv n/s$,
with $s$ the total entropy density in the SM and DS. For definiteness we choose a Majorana DM
particle with $m_\chi=100\,\mathrm{GeV}$ and a constant annihilation {\it amplitude}
(for which we provide a closed expression for $\mathfrak{C}$ in Appendix B)
that would result in the correct relic density in the standard treatment
(translating to a value of $\langle \sigma v\rangle_{T_\chi \to0}$ about $10\%$ larger than the orange
lines in Fig.~\ref{fig:no_mu_dark}).
The red curves show the case of $m_S = 0$ for which, following the discussion above, we explicitly set $\mu_S = 0$.
The resulting evolution of $\chi$ (red solid line) therefore coincides exactly with the result of the standard
treatment of solving Eq.~\eqref{boltzn_standard}. We note that the increase in $Y_S$ around $T_\chi \sim m_\chi$
is due to the Boltzmann suppression of $\chi$, analogous to the increase in $n_\gamma / s$ 
during $e^+ e^-$ annihilation in the SM.

For more degenerate masses 
(green and purple lines in Fig.~\ref{fig:Yx}), we allow all chemical potentials to evolve freely.
This leads to a rise in $\mu_S$, compensating the would-be Boltzmann suppression of $S$, and an asymptotic abundance
$Y_S^{\rm final}\approx Y_S^{\rm initial}+Y_\chi^{\rm initial}$ because $Y_\chi^{\rm initial}\gg Y_\chi^{\rm final}$.
The greater number of $S$ particles then delays the Boltzmann suppression of $n_\chi$ from around $T_\chi \sim m_\chi$ 
to when the mean kinetic energy of $S$ drops below $\delta m$, roughly around $T_\chi \sim \delta m$. For reference we 
also show an application of Eq.~\eqref{boltzn_standard} (dotted lines) assuming thermal equilibrium of $S$ 
with additional 
massless DS heat bath particles such that $\mu_S = 0$ and $T_\chi \propto a^{-1}$ with the scale factor $a$.
Comparing the purple lines $(\delta m/m_\chi = 10^{-2})$, {\it e.g.}, Boltzmann suppression of $\chi$ for the solid line occurs
at temperatures $T_\chi$ around two orders of magnitude smaller than for the dotted line, or $a$
one order of magnitude larger ($T_\chi \propto a^{-2}$ at $T_\chi \lesssim m_S$ for the solid line). Approximating the annihilation rate by
$\langle \sigma v \rangle n_\chi \propto a^{-3}$, whereas the dilution by cosmic expansion is $H \propto a^{-2}$,
this implies that freeze-out happens when $\chi$ is less Boltzmann-suppressed and $Y_\chi$ is enhanced by $\sim a$, i.e.\
around one order of magnitude. In general, the correct treatment of the chemical potentials thus leads to an enhanced DM
abundance compared to the `na\"ive' assumption of $\mu_S = 0$ and $T_\chi \propto a^{-1}$.
Comparing instead to the $m_S=0$ case, cf.~the standard situation depicted in
Fig.~\ref{fig:no_mu_dark}, $Y_\chi^{\rm final}$ first decreases up to a mass ratio of
$m_S/m_\chi=0.4$ (green lines) as a larger $m_S$ implies a faster decrease of $T_\chi$ with time around $T_\chi \sim m_\chi$ such that, in fact, the SM temperature $T$ is somewhat \emph{larger} around freeze-out. Approximating the annihilation rate as above, freeze-out occurs when $H/s \sim \langle \sigma v \rangle n_\chi / s \propto 1/T$ (the SM dominates the energy and entropy densities), leading to a slight decrease in $Y_\chi = n_\chi / s$. For even larger $m_S$, this effect is compensated by the delayed Boltzmann suppression of $\chi$ and the DM abundance increases as $S$ and $\chi$ become more and more degenerate.

\begin{figure}[t]
\includegraphics[width=0.95\columnwidth]{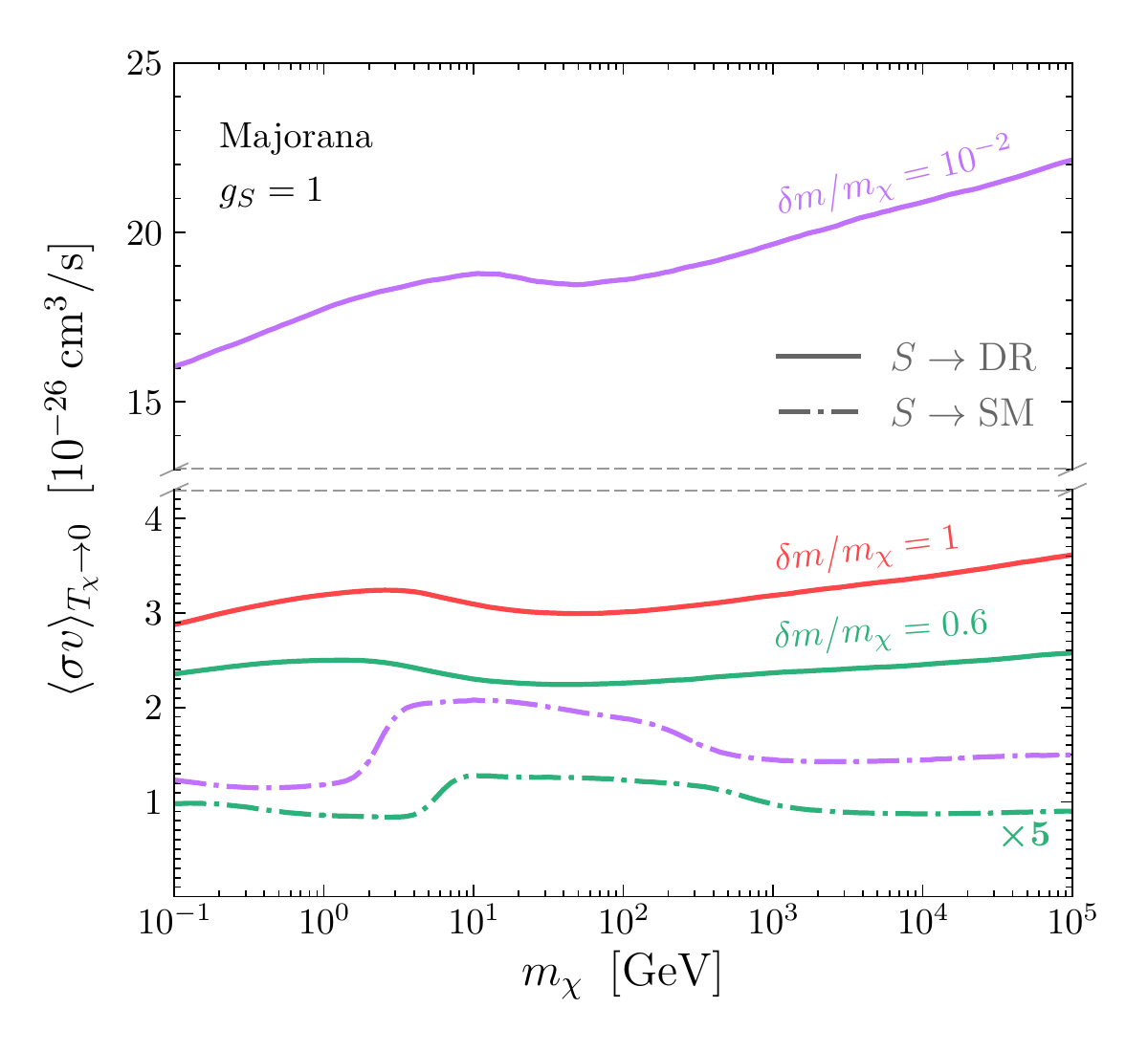}
\caption{The required value of the thermally averaged annihilation cross-section, $\langle\sigma v\rangle_{T_\chi \to 0}$, that
results in a relic density of Majorana DM particles with a constant
$\left|\overbar{\mathcal{M}}_{\chi\bar{\chi}\to SS}\right|^2$ matching the observed DM abundance.
Colors correspond to the same mass ratios as in Fig.~\ref{fig:Yx}, while the line style distinguishes whether $S$ decays into dark radiation (solid, independent of lifetime $\tau_S$) or into SM states
(dash-dotted, for $\tau_S = 1 \, \mathrm{s} \times (1 \, \mathrm{GeV} / m_S)^2$).
}
\label{fig:rd_mudark}
\end{figure}

For $S$ close in mass to $\chi$, the final DM relic abundance will not only depend on the decoupling process but also on how $S$
decays after freeze-out.
If $S$ was stable, in particular, it would simply contribute to the total DM density, by far overshooting the observed value
(unless allowing for sufficiently small temperature ratios
$\xi_{T \rightarrow \infty} \ll 1$, thus relaxing our assumption of initial thermal contact between SM and DS). In the following we assume a lifetime $\tau_S$ of $S$ such that the decays occur only after freeze-out, implying a negligible impact on the freeze-out process itself.
In Fig.~\ref{fig:rd_mudark} we explore two concrete decay scenarios, by showing the `thermal' annihilation cross-section for 
the same mass ratios as discussed in Fig.~\ref{fig:Yx}.\footnote{%
This is implemented by adding $-n_S/\tau_S$ to the r.h.s.\ of Eq.~\eqref{eq:boltzmann_similar} for $n_S$,
$-m_S n_S / \tau_S$ to the r.h.s.\ of Eq.~\eqref{eq:friedmann_similar}, and an additional energy density in dark radiation
$\dot{\rho}_\mathrm{DR} + 4 H \rho_\mathrm{DR} = m_S n_S / \tau_S$ for decays in effectively massless DS states, or
$\dot{\rho}_\mathrm{SM} + 3 H(\rho_\mathrm{SM} + P_\mathrm{SM}) = m_S n_S / \tau_S$ for decays into SM particles.} 
The first scenario is $S$ decaying to effectively massless DS states, 
or dark radiation (DR), and indicated by solid lines. The additional effective relativistic d.o.f.~resulting from the decay of $S$ 
will in general depend on the lifetime $\tau_S$, because the energy densities of matter and radiation red-shift differently.
As already for $m_S = 0$ one has $\Delta N_\mathrm{eff} = 0.104$ (see above), the case $m_S \sim m_\chi$ is generally in 
conflict with the CMB limit even if the decay happens shortly after freeze-out. The second example
(dash-dotted lines) considers $S$ decays to SM states. In this case,  the resulting entropy injection into the SM plasma will 
lead to a dilution of DM, lowering the required DM annihilation cross-section. 
This effect has recently been argued to allow for DM masses above the na\"ive
unitarity limit~\cite{Berlin:2016vnh,Berlin:2016gtr,Cirelli:2018iax}. Note that the lifetime 
$\tau_S = 1 \, \mathrm{s} \times (1 \, \mathrm{GeV} / m_S)^2$ chosen here for illustration is expected to be in conflict with 
observations of primordial 
element abundances for $\tau_S > 0.1 \, \mathrm{s}$~\cite{Hufnagel:2018bjp}, i.e.\ $m_S \lesssim 3 \, \mathrm{GeV}$.

To summarize, the solid lines in Fig.~\ref{fig:rd_mudark} show the required DM annihilation cross-section to obtain the 
observed DM abundance assuming $S$ decays \emph{without} injecting entropy in the SM and thus diluting the DM 
abundance. These lines therefore provide an \emph{upper} limit to scenarios where $S$ decays into the SM after DM freeze-out, 
as exemplary illustrated by the dash-dotted lines.
It is evident that the required DM annihilation cross-section can be very different from the canonical value shown in 
Fig.~\ref{fig:no_mu_dark}, in particular for small mass differences.
In the extreme case of degenerate masses, $m_S = m_\chi$, no Boltzmann suppression of $\chi$ can occur --
independently of the DM annihilation cross-section -- implying that the observed DM abundance can only be achieved for
sufficiently small temperature ratios $\xi_{T \rightarrow \infty} \ll 1$ as discussed above for a stable $S$.

\smallskip
\paragraph*{Discussion.---}%
For the choice of parameters discussed above we explicitly checked (see Appendix B)
that the assumption of
kinetic equilibrium is always satisfied during the freeze-out process, justifying our ansatz for the phase-space
distributions $f_a$. 
Let us stress that this is particularly important for small mass splittings, where $\mu_S \simeq m_S$
makes it mandatory to  include the full quantum statistics for all particles when aiming for precision
calculations of the relic density. The commonly used assumption
of a Maxwell-Boltzmann distribution is, in other words, no longer justified.
For the green lines in Fig.~\ref{fig:rd_mudark}, for example, with $m_S=0.4\,m_\chi$,
ignoring the effect of quantum statistics leads to an {\it over}estimate of the final DM abundance by 
about $3$\,\% for hidden sector decay (solid line) and more than $10$\,\% for decay into SM particles 
(dash-dotted lines). 
While we leave a more detailed investigation for future work, let us stress that the effect is typically larger 
{\it during} the freeze-out process, at higher temperatures, and hence potentially more relevant for 
(semi-)relativistic freeze-out;
for hidden sector decay, furthermore, the impact is also larger on the abundance of $S$, 
which is very sensitive to
ever more stringent constraints from $\Delta N_{\rm eff}$.

So far we have focussed on a fully secluded DS, in which case the most prominent observables to test such models
are $\Omega_{\rm DM}$ and $\Delta N_{\rm eff}$. It is however worth mentioning that in many models there are
additional tiny couplings to the SM that would allow further experimental signatures. A setup where hidden sector freeze-out
can naturally occur while still allowing for sufficiently large couplings to the SM to be probed by particle physics experiments,
{\it e.g.}, are scalar or pseudoscalar mediators with Yukawa-like coupling structure~\cite{Dolan:2014ska,Alekhin:2015byh,Kahlhoefer:2017umn,Evans:2017kti,Winkler:2018qyg,Bondarenko:2019vrb}.
Also indirect DM searches for secluded dark sectors~\cite{Elor:2015bho} provide a potentially promising avenue,
in particular for the strongly enhanced annihilation rates necessary to accommodate DM degenerate in mass with
its annihilation products.

\smallskip
\paragraph*{Conclusions.---}%
In this work we have presented a framework for precision calculations of DM freeze-out in a secluded sector, 
matching the observational accuracy on the one hand, and the increasing demand for consistent interpretations
of phenomenological dark sector studies on the other hand.
We have provided new benchmark `thermal' annihilation cross-sections for relativistic heat bath particles,
and demonstrated that the difference to the standard treatment can be even larger for non-relativistic
DM annihilation products. The latter case is intrinsically strongly model-dependent, and will be studied in
more detail elsewhere. 
Further interesting extensions, not the least in view of the significant model-building
activity in these areas, would be to generalize the precision relic calculations presented here to models where
the DM particles in the hidden sector do
not obey a $Z_2$ symmetry~\cite{Carlson:1992fn,Hochberg:2014dra,Bernal:2015ova}, are 
asymmetric~\cite{Kaplan:2009ag} or have a relic abundance set by freeze-in rather than 
freeze-out~\cite{Hall:2009bx,Hall:2010jx}.

\vfill
 \paragraph*{Acknowledgements.---}%
This work is supported by the ERC Starting Grant
‘NewAve’ (638528) as well as by the Deutsche Forschungsgemeinschaft under Germany’s
Excellence Strategy – EXC 2121 ‘Quantum Universe’ – 390833306.
\appendix

\section{A. DM annihilation via $p$-wave}
\label{app_pwave}

In the case of $s$-wave annihilation to massless final states, the velocity-weighted annihilation cross-section is
constant in the limit of small DM velocities, resulting in $\langle\sigma v\rangle=\sigma v_{\rm lab}$.
This simplified ansatz for $\langle\sigma v\rangle$ (neglecting higher-order contributions in $v$,
following common practice)
has been presented in Fig.~\ref{fig:no_mu_dark} in the main text, both for DM annihilating to SM
particles and for situations in which the relic abundance is set via freeze-out in a hidden sector.

\begin{figure}[t]
\includegraphics[width=0.95\columnwidth]{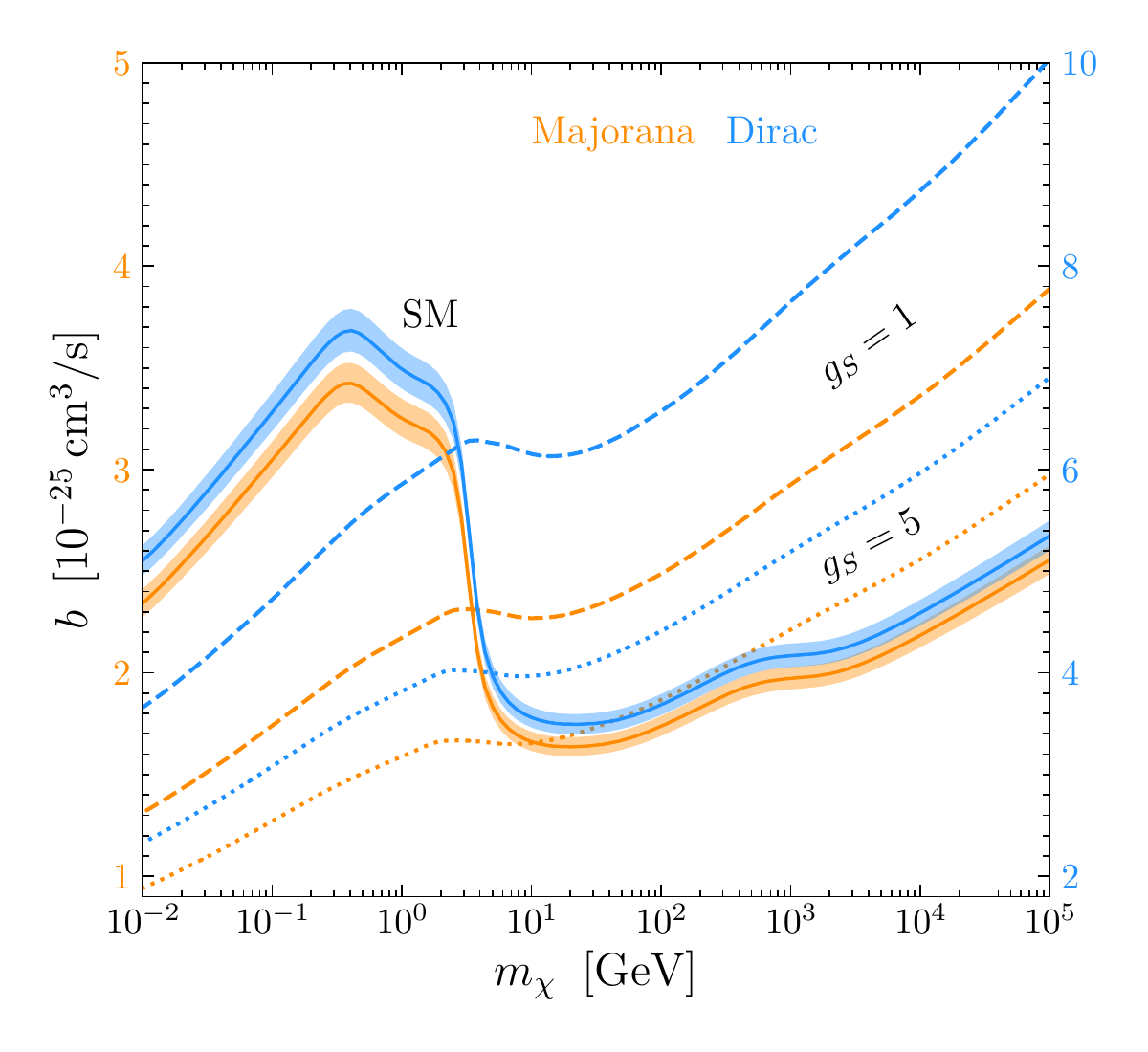}
\caption{Same as Fig.~\ref{fig:no_mu_dark} in the main text, but for $p$-wave annihilation
with $\sigma v_{\rm lab}=b\,v_{\rm lab}^2$.
}
\label{fig:no_mu_dark_pwave}
\end{figure}

Here we complement this by considering instead the case of $p$-wave annihilation, which also has been
frequently considered for DS freeze-out production of DM~\cite{Tulin:2013teo,Kaplinghat:2013yxa,Dolan:2014ska,Alekhin:2015byh,Kahlhoefer:2017umn,Evans:2017kti,Winkler:2018qyg}.
To describe such models, we will again take
a simplified ansatz for the cross-section by only keeping the leading term in the DM velocities,
\be
 \sigma v_{\rm lab}=b\,v_{\rm lab}^2\,,
\ee
where we assume $b$ to be constant.
For the thermally averaged cross-section entering in the Boltzmann equation, Eq.~\eqref{boltzn_standard} in the main text,
this implies $\langle\sigma v \rangle=b\times\left[6 (x/\xi)^{-1} -27 (x/\xi)^{-2}  +...\right]$.
The value of $b$ resulting in the correct DM relic abundance in this case is shown in
Fig.~\ref{fig:no_mu_dark_pwave}, for the same choice of DM models (Dirac and Majorana
fermions, respectively) and heat bath components as in Fig.~\ref{fig:no_mu_dark} in the main text.

In comparison, the main differences in these figures are that {\it i)}  the value of $b$ resulting in the
correct relic density is about one order of magnitude larger than the value of $\langle\sigma v\rangle$
required in the case of $s$-wave annihilation and that {\it ii)} this `thermal' value of $b$ rises faster
with $m_\chi$ than its $s$-wave counterpart. Both of this can be traced back to the fact that also for $p$-wave
annihilation it is $\langle\sigma v\rangle$ around chemical decoupling, and not $b$, that sets the relic
density. In the SM case, e.g., $b/\langle\sigma v\rangle\approx x_{\rm cd}/6$,
where $x_{\rm cd}$ depends logarithmically on the DM mass and rises from $x_{\rm cd}\approx18.8$
(for $m_\chi=100$\,MeV) to $x_{\rm cd}\approx31.6$ (for $m_\chi=100$\,TeV).
The above estimate should be corrected by another factor of about 2  because
decoupling does not happen instantaneously, and
$\int\!\di T \langle\sigma v\rangle^{p-{\rm wave}}/\int\!\di T \langle\sigma v\rangle^{s-{\rm wave}}\approx 1/2$~(as
first stressed in Ref.~\cite{Scherrer:1985zt}).
The same general trend, finally, is also visible for annihilations in the hidden sector, with $\xi\neq1$.
Compared to Fig.~\ref{fig:no_mu_dark} in the main text, furthermore, the difference between SM and DS results is
somewhat larger because $\xi$ enters directly in $\langle\sigma v\rangle$.

\section{B. Collision term including chemical potentials}

For general two-body annihilation processes $\chi\bar\chi\leftrightarrow SS'$, and assuming $CP$-invariance, the
integrated collision operator from Eq.~\eqref{eq:boltzmann_similar} in the main text takes the form
\begin{align}
\mathfrak{C} & = g_\chi^2 \int (2\pi)^4\delta(p_\chi + p_{\bar{\chi}} - p_S - p_{S'})|\mathcal{\overbar{M}}_{\chi\bar{\chi}\rightarrow SS'}|^2\nonumber \\
& \times \big[ f_{S}f_{S'} (1-f_\chi)(1-f_{\bar{\chi}}) - f_\chi f_{\bar{\chi}} (1+f_S)(1+f_{S'}) \big] \nonumber \\[0.2cm]
& \times \di \Pi_\chi \di \Pi_{\bar{\chi}} \di \Pi_S \di \Pi_{S'}\,,
\label{eq:cfull}
\end{align}
where $\di \Pi_a = \di^3 p_a/(2\pi)^3 2E_a$, integration is implied over \textit{all} (not only physically distinct) momentum configurations, and $|\mathcal{\overbar{M}}_{\chi\bar{\chi}\rightarrow SS'}|^2$ is the
squared matrix element, averaged (summed) over the spins and other internal degrees of freedom of all initial (final) state particles.
We assume all involved particles to be in kinetic
equilibrium, i.e.~the phase-space distributions take the form $f_a=1/[e^{(E_a-\mu_a)/T_\chi}\pm1]$, with
$a \in \{ \chi, \bar\chi, S, S'\}$ and the $-$ ($+$) sign is used for bosons (fermions). In the special case of
a constant matrix element -- which is justified for contact-like interactions and which we adopt as benchmark scenario
in the main text -- Eq.~\eqref{eq:cfull} can be simplified to
\begin{align}
\mathfrak{C} = \frac{g_\chi^2|\mathcal{\overbar{M}}_{\chi\bar{\chi}\rightarrow SS}|^2}{512\pi^5} \int_{m_\chi}^\infty \int_{m_\chi}^\infty \int_{-1}^1 p_\chi p_{\bar{\chi}} \mathcal{K}\;\di\cos\theta \di E_\chi \di E_{\bar{\chi}}\,.
\end{align}
Moreover,
\begin{align}
\mathcal{K} \equiv \alpha_* (1-f_\chi)(1-f_{\bar{\chi}}) - f_\chi f_{\bar{\chi}}\left(\beta+2\alpha + \alpha_{*}\right)\,,
\end{align}
with
\begin{align}
& \beta \;\, \equiv \;\, \sqrt{1-\frac{4m_S^2}{E'^2-p'^2}}\,,\\
& \alpha \;\, \equiv \;\, \frac{T_\chi}{p'} \log\left[ \frac{e^{E'/T_\chi} - e^{(E'-p'\beta+2\mu_S)/(2T_\chi)}}{e^{E'/T_\chi} - e^{(E'+p'\beta + 2\mu_S)/(2T_\chi)}} \right]\,,\\
& \alpha_* \equiv \;\, \frac{\beta +  2\alpha}{e^{(E'-2\mu_S)/T_\chi}-1}\,,
\end{align}
where
$p' \equiv |\vec{p}_\chi + \vec{p}_{\bar{\chi}}| = (p_\chi^2 + p_{\bar{\chi}}^2 + 2p_\chi p_{\bar{\chi}}\cos\theta)^{1/2}$
and $E' \equiv E_\chi + E_{\bar{\chi}}$.

For highly non-relativistic DM, the annihilation cross-section for a constant matrix element becomes
independent of the center-of-mass energy, and hence $\sigma v_{\rm lab}\simeq \langle \sigma v \rangle$.
In this limit, annihilation cross-section and amplitude are related as
\begin{align}
|\mathcal{\overbar{M}}_{\chi\bar{\chi}\rightarrow SS}|^2 \simeq \frac{64 \pi m_\chi^2}{\sqrt{1-m_S^2/m_\chi^2}}
 \langle \sigma_{\chi\bar{\chi} \rightarrow SS} v \rangle_{T_\chi \rightarrow 0}
\,.
\end{align}
In the simplest models, the same constant matrix element also describes the scattering process
$\chi S\leftrightarrow \chi S$,
in which case the above expression provides a convenient means of estimating the time of kinetic decoupling for a
given value of
$\langle \sigma v \rangle$. For $m_\chi\sim m_S$, e.g., this happens when the scattering rate falls behind
the Hubble rate, $n_S\langle \sigma_{\chi S\leftrightarrow \chi S} v \rangle\sim H$, while for $m_\chi\gg m_S$
it is instead the (smaller) momentum exchange rate $\gamma$ that provides the relevant scale (see, e.g.,
Refs.~\cite{Bringmann:2009vf,Binder:2017rgn}). Using this condition, we explicitly checked that $S$ and $\chi$ remain in kinetic equilibrium during the freeze-out process.

\bibliography{freezeout.bib}

\end{document}